\documentclass[amsmath,amsfonts,amssymb,twocolumn,showpacs]{revtex4}
\usepackage{graphicx,psfrag,color}% Include figure files
\usepackage{dcolumn}% Align table columns on decimal point
\usepackage{bm}% bold math
\usepackage{mathbbol}

\newcommand{\bk}{{\bf k}}

\begin{document}

\title{The Cooper pair from radio-frequency excitations in ultracold gases}

\author{J.~Dukelsky{$^1$} and G. Ortiz{$^2$}}
\address{$^{1}$ Instituto de Estructura de la Materia, CSIC, Serrano
123, 28006 Madrid, Spain \\
$^{2}$ Department of Physics, Indiana University,
Bloomington, IN 47405, USA}

\begin{abstract}
We discuss the concept of Cooper pair in the context of recent
experimental studies of radio-frequency excitations in ultracold atomic
gases. We argue that the threshold energy determines the size of the
Cooper pair  emergent from the exact solution of the reduced BCS
problem, and  elaborate on the physical distinction between bosonic and
fermionic Cooper pairs.
\end{abstract}

\maketitle

A few years ago \cite{ours1} we thoroughly analyzed the notion and
nature of Cooper pairs in a correlated Fermi system. As shown by Cooper,
an isolated pair in an inert Fermi sea constitutes a  true bound state
while a correlated pair in a superconductor or superfluid medium behaves
as a quasi-resonance. Simply said, the single pair case corresponds to a
two-particle problem while the latter represents a true many-body
problem. We also raised the provocative question {\it What is a Cooper
pair?}, and argued that the Cooper pair in a correlated Fermi superfluid
is a statistical concept that acquires physical significance only
through the particular experimental probe one uses to test the system.
For example, if one measures the antiparallel spin part of the static
density-density correlation function of the Fermi system, one is probing
the (Fourier transformed) pair wavefunction $u_\bk v_\bk$.

Recent experimental studies of radio-frequency (RF) excitations in
ultracold Fermi gases do not measure the latter pair wavefunction.
Indeed, by using a mean-field BCS description of the system one may
conclude that they do probe the BCS pair wavefunction $v_\bk/u_\bk$.
But, what property of this pair wavefunction do they really probe? A
main goal of this short Note is to argue that  what these experiments
really probe are the properties of the Cooper pair concept that emerges
from the exact solution of the reduced BCS problem. For example, we show
that the pair size derived from the threshold energy as recently
proposed in Ref. \cite{Kett} is nothing more than the notion of pair
size defined in Ref. \cite{ours1}.

The reduced s-wave paring BCS Hamiltonian, for fermions enclosed in a
volume $V$ and with non-interacting dispersion
$\varepsilon_{\mathbf{k}}$, can be written as
\begin{eqnarray}
H &=&\!\!\sum_{\mathbf{k}}\varepsilon_{\mathbf{k}}\
{n}_{\mathbf{k}}+\frac{G}{V}\sum_{\mathbf{ k,k^{\prime }}}\
c_{\mathbf{k\uparrow }}^{\dagger }c_{-\mathbf{k\downarrow }
}^{\dagger }c_{-\mathbf{k^{\prime }\downarrow
}}^{\;}c_{\mathbf{k^{\prime }\uparrow }}^{\;}  \ ,
\label{BCS}
\end{eqnarray}
which includes all terms with time-reversed pairs
($\bk\uparrow,-\bk\downarrow$) from a contact interaction between
fermions of mass $m$, momentum $\mathbf{k}$, and spin component
$\sigma=\uparrow,\downarrow$ (represented by the (creation) operator
$c_{\mathbf{k\sigma}}^{\dagger }$). In first quantization language, the
resulting {\it exact} BCS state  \cite{Rich} for $M=N/2$ pairs is given
by ($x_j=({\bf r}_j, \sigma_j)$)
\begin{eqnarray}
\Psi(x_1,\cdots,x_N)=  {\cal A} \left [ \phi_1(x_1, x_2) \cdots
\phi_{M}(x_{N-1}, x_N) \right ] \ , \label{Var}
\end{eqnarray}
with ${\cal A}$ the antisymmetrizer, and the {\it pair} state
\begin{equation}
%\phi_{n}(\mathbf{r}_1\sigma_1,\mathbf{r}_2\sigma_2)=\frac{1}{\sqrt{V}}
\phi_{\alpha}(x_i,x_j)=\
e^{i\mathbf{q}\cdot\frac{(\mathbf{r}_i+\mathbf{r}_j)}{2}} \
\varphi_\alpha(\mathbf{r}_i-\mathbf{r}_j)  \ \chi(\sigma_i,\sigma_j) \ ,
\label{Var2}
\end{equation}
where $\chi$ is the spin function, $\mathbf{q}$ is the pair center-of-mass momentum, and
$\varphi_\alpha(\mathbf{r})$ the internal (relative coordinate) wavefunction. The many-body wavefunction
(\ref{Var}) describes a collection of $M$ different pairs ({\ref{Var2}) which, depending upon the strength of the
interaction between particles $G$, may represent either a quasimolecular resonant state or a scattering state
(i.e., free fermions). The state (\ref{Var}) is the natural generalization of the Cooper-pair problem without an
inert Fermi sea, and with all pairs subjected to the pairing interaction. Only in the extreme BEC limit this
physical state and Leggett's ansatz \cite{leggett}, which assumes {\it all equal pairs}, are equivalent.

For a uniform 3$D$ system in the thermodynamic limit the exact solution of the Hamiltonian (\ref{BCS}) reduces to
the BCS gap equation \cite{gaudin}. The singularity displayed by this equation can be regularized by introducing
the scattering length $a_s$, resulting in a model that describes the BCS-to-BEC crossover in terms of a single
parameter $\eta=1/k_F a_s$, with $k_F$ the Fermi momentum \cite{leggett}. We studied the character and nature of
these pairs in the BCS-to-BEC crossover region. Our results \cite{ours1} suggested a scenario in which correlated
pairs and free fermions coexist in the BCS region while there is a collection of quasibound molecules in the BEC
region. More precisely, the pair wavefunction has the {\it universal} form \cite{ours1}
\begin{eqnarray}
\varphi_E\left(  r\right)  =\sqrt{\frac{m \operatorname{Im}\left(  \sqrt
{E}\right)  }{2\pi \hbar^2}}\frac{e^{-r\sqrt{-mE/\hbar^2}}}{r} ,
\label{Cooperpair}
\end{eqnarray}
independently of the value of $\eta$, where $E$ is the spectral parameter or the pair energy determined from the
Richardson (Bethe ansatz) equations \cite{Rich}. It is important to emphasize that the many-body state (\ref{Var})
with the pair wavefunctions given by Eq. (\ref{Cooperpair}) is not an ansatz but it is the state that results from
the exact solution to Eq. (\ref{BCS}). The character of the pair wavefunction is determined by $E$ which is, in
general, a complex number. If $E$ is real and positive then $\varphi_E$ represents a scattering state; if it is
complex with a positive real part it is a Cooper resonance, while if it is complex with a negative real part it is
a quasibound molecular state. As a function of $\eta$, the distribution of pairs in the correlated state
$\Psi(x_1,\cdots,x_N)$ may have a different character, suggesting the following identifications

- {\bf BCS regime:} a mixture of real $E$, and complex $E$ with $\operatorname{Re}(E) > 0$ (Cooper resonances and
free fermions).

- {\bf Pseudogap regime:} a mixture of complex $E$ with both $\operatorname{Re}(E) \gtrless 0$ (Cooper resonances
and quasimolecules).

- {\bf BEC regime:} complex $E$ with $\operatorname{Re}(E) < 0$ (quasimolecules).

The size of each pair, defined as the mean square radius
$\xi_E=\sqrt{\langle \varphi_E |  r^2 | \varphi_E \rangle}$, is given by
\begin{eqnarray}
\xi_{E}^{2}=\frac{\hbar^2}{2 m\left(  \operatorname{Im}\sqrt{E}\right)
^{2}} =\frac{\hslash^{2}}{2mE_{p}}
\end{eqnarray}
where
\begin{eqnarray}
2E_{p}=\sqrt{\operatorname{Re}\left(E\right)^{2}+\operatorname{Im}\left(
E\right)^{2}}- \operatorname{Re}\left(E\right)
\end{eqnarray}

For a given value of the coupling $\eta$, the most correlated pairs are located close to the two extremes of a
complex arc $\Gamma$ with values $E=2(\mu\pm i\Delta)$ (see Ref. \cite{ours1,gaudin}). Inserting this value in the
equation above leads to
\begin{eqnarray}
E_{p}=\sqrt{\mu^{2}+\Delta^{2}}-\mu , \label{Thresh}
\end{eqnarray}

In what follows we present results for some particular values of $\eta$
\begin{eqnarray}
E_{p}=\begin{cases}
\frac{\Delta^2}{2 \epsilon_F} & \mbox{ when } \eta \rightarrow - \infty
\mbox{ (BCS)} \cr\cr
\frac{(1+x) \ \epsilon_F }{E\left( \frac{1-x}{2}\right)^{\frac{2}{3}}} &
\mbox{ when } \eta=0 \mbox{ (resonance)} \cr\cr
\sqrt{2\eta} \ \epsilon_F& \mbox{ when }
\eta=\frac{8\pi^{2/3}}{\Gamma[\frac{1}{4}]^{8/3}}\mbox{ ($\mu=0$)} \cr\cr
\frac{\hbar^2}{m a_s^2} & \mbox{ when } \eta \rightarrow \infty \mbox{
(BEC)}\cr\cr
\end{cases}
\end{eqnarray}
where $x \sim -0.65222953$ is the root of the Legendre function of the
first kind $P_{\frac{1}{2}}$, i.e. $P_{\frac{1}{2}}(x)=0$, and $E(y)$ is
the complete elliptic integral of the second kind ($(1+x)/E\left(
\frac{1-x}{2}\right)^{\frac{2}{3}}\approx 0.314912$). Figure \ref{fig1}
compares the size of this correlated pair with the sizes of the pair
correlation function $\xi_P$, and the BCS wavefunction $\xi_{BCS}$. As
already mentioned in \cite{ours1} and recently suggested in \cite{Kett},
the size of the Cooper pair in the BCS region is {\em twice} the size of
the pair correlation function. Moreover, while the size of the pair
correlation function crosses the interparticle distance at $\eta \sim
-0.5$ (where the fraction of the condensate is $\sim 0.5$), the Cooper
pair size crosses the interparticle distance at resonance (where the
fraction of the condensate approaches $1$). The three wavefunctions
coincide in the extreme BEC limit.

\begin{figure}[htb]
\vspace*{-0.3cm}
\hspace*{-0.5cm}
\includegraphics[angle=0,width=9.6cm,scale=1.0]{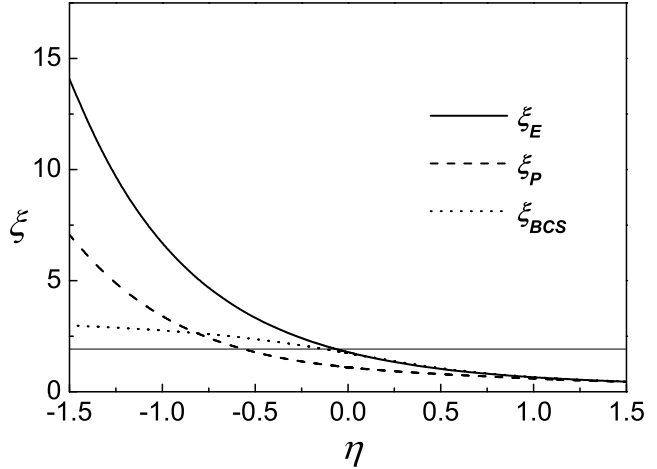}
\caption{Size of the Cooper pair $\xi_E$ derived from the exact solution
of the reduced BCS Hamiltonian of Eq. (\ref{BCS}) in units of $1/k_F$.
We have plotted sizes $\xi_E$, $\xi_P$, and $\xi_{BCS}$ for the most
correlated pair which corresponds to $E=2(\mu\pm i\Delta)$ (solid line),
the pair correlation function $u_\bk v_\bk$ (dashed line), and the BCS
wavefunction $v_\bk/u_\bk$ (dotted line), respectively. The thin solid
line is the interparticle distance $r_s=\sqrt[3]{9\pi/4}$. }
\label{fig1}
\end{figure}

%By analogy to resonance phenomena one can interpret
%$\tau_E=\sqrt{2}/\operatorname{Re}(\sqrt{E})$ as the lifetime of
%the Cooper pair. Indeed, $\tau_E$ monotonically increases from the
%BCS to the BEC region and $\tau_E\rightarrow \infty$ when $\eta \rightarrow
%+ \infty$, corresponding to tightly bound molecules. The lifetime (we
%need to write in teh same units as above, here I am using the same units
%as in our paper) for the most correlated pairs is given by
%\begin{eqnarray}
%\tau_{E}=\frac{\sqrt{2}}{\sqrt{\mu+\sqrt{\mu^{2}+\Delta^{2}}}} ,
%\end{eqnarray}

%One can provide an operational physical interpretation to those pair
%wavefunctions in the case of ultracold atomic gases. The hyperfine
%structure of those atoms allows for RF transitions to a third level with
%the end result that the structure of the Cooper pair we defined above
%\cite{ours1} can be probed. Following steps similar to those of  Ref.
%\cite{Kett1} but simply adapted to deal with the exact solution of Eq.
%(\ref{BCS}) one defines an RF operator
%\begin{equation}
%\hat{V}=V_0 \sum_\bk c^\dagger_{\bk 3} c^{\;}_{\bk \uparrow} + {\rm
%h.c.} ,
%\end{equation} and show that the transition rate at which particles leave the exact BCS state is proportional to
%the pair wavefunction...

It is important to remark that the most correlated pair determines the
minimum energy or threshold energy to excite a particle in a
RF-excitation experiment:  $E_{\sf th}=E_p$ in (\ref{Thresh}). In this
respect, the operative definition  $\xi _{th}^2=\hbar^2 /2 m E_{th}$
used in \cite{Kett} to extract the size of the Cooper pair from a
fitting of the threshold energy represents a direct observation of the
Cooper pair wavefunction (\ref{Cooperpair}) at the extremes of the arc
for each value of $\eta$. $E_{\sf th}$ does not provide information
about the size of either the pair wavefunctions $v_\bk/u_\bk$ or  $u_\bk
v_\bk$.

% In this way, RF-excitation
%experiments provide a natural way to probe properties of the Cooper pair defined in Eq. (\ref{Cooperpair}). It is
%interesting to remark that the most correlated pair determines the threshold energy $E_{\sf th}=E_p$ of the
%RF-excitation spectrum. Obviously, $E_{\sf th}$ does not provide information about the size of either the pair
%$v_\bk/u_\bk$ or  $u_\bk v_\bk$.

Finally, we would like to mention an essential difference between the
structure of fermionic and bosonic atomic pairs. It has been shown that
the Hamiltonian of Eq. (\ref{BCS}) is exactly solvable for fermionic as
well as for bosonic systems \cite{bos}. In the bosonic case the exact
eigenstate is determined from the same Eqs.
(\ref{Var})-(\ref{Cooperpair}) with a symmetrizer in (\ref{Var}) instead
of an antisymmetrizer. However, the bosonic ground state has always real
and negative spectral parameters $E$ for an arbitrary attractive
interaction $G$. Physically, it represents a collection of {\it bound}
diatomic molecules described by the Cooper pair wavefunction
(\ref{Cooperpair}) in different excited states.

In summary, in this short Note we argued that RF excitation experiments
probe the notion of Cooper pair emergent from the exact solution of the
reduced BCS problem. We emphasize that the Cooper pair wavefunction
(\ref{Cooperpair}) has a {\it universal form} as the  RF dissociation
spectrum in the whole crossover. In particular, the size derived from
the threshold energy is directly related to the size of the Cooper pair
defined in Ref. \cite{ours1}.

%Imagine we have a certain number of trapped atoms forming
%a condensate at sufficiently low temperatures. We want to open the trap
%quickly and ask the question: what are the pairs that remain in the
%center of the device? The answer depends upon the character of the pair
%forming the condensate. Unless the pair has a complex pair energy with a
%negative real part, it will always prefer to leave the trap. This simple
%characterization provides a very powerful prediction after one opens the
%trap: If the gas is in the BCS regime there will be no pairs in the
%center of the trap; if it is in the BEC regime all pairs will remain;
%while in the interesting Pseudogap regime some pairs will stay while
%others will fly appart.


\begin{thebibliography}{28}

%\bibitem{kitaev}
%A. Kitaev, Ann. Phys. {\bf 303}, 2 (2003).

\bibitem{ours1}
G. Ortiz and J. Dukelsky, Phys. Rev. A {\bf 72}, 043611 (2005); G. Ortiz
and J. Dukelsky, {\it Condensed Matter Theories}, Vol. 21, Eds. H. Akai,
A. Hosaka, H. Toki, and F. B. Malik (Nova Science Publishers, Inc.,
Huntington,  New York, (2007)).

\bibitem{Kett}
Ch. H. Schunck {\it et al.}, arXiv:0802.0341.

\bibitem{Rich}
R. W. Richardson, Phys. Lett. {\bf 3}, 277 (1963); Nucl. Phys. {\bf 52},
221 (1964).

\bibitem{leggett}
A. J. Leggett, in {\it Modern trends in the theory of condensed matter},
edited by A. Pekalski and R. Przystawa (Springer Verlag, Berlin, 1980);
D. M. Eagles, Phys. Rev. {\bf 186}, 456 (1969).


\bibitem{gaudin} M. Gaudin, {\it Mod\`eles exactement r\'esolus}
(Les Editions de Physique, Les Ulis, 1995), p. 261; J. M. Roman, G. Sierra, and J. Dukelsky, Nucl. Phys. B {\bf
634}, 483 (2002).


\bibitem{Kett1}
W. Ketterle and  M. W. Zwierlein, arXiv:0801.2500.

\bibitem{bos}
R. W. Richardson, J. Math. Phys. (NY), {\bf 9}, 1327 (1968);
J. Dukelsky and P. Schuck, Phys. Rev. Lett. {\bf 86}, 4207 (2001).

\end{thebibliography}
\end{document}